\documentclass[%
reprint,
nofootinbib,
amsmath,amssymb,
aps,
floatfix,
10pt
]{revtex4-2} 
\usepackage[latin1]{inputenc}
\usepackage{amsmath,amssymb}
\usepackage{mathrsfs} 
\usepackage[capitalise]{cleveref}
\usepackage{siunitx}
\usepackage{braket}
\usepackage{calc}
\usepackage{tabularx}
\usepackage{dsfont}
\usepackage{color}
\usepackage{seqsplit}
\usepackage{ifthen}
\usepackage[normalem]{ulem}
\usepackage{graphicx}
\usepackage{mathtools} 

\allowdisplaybreaks
\newcommand{\Eins}{\mathds{1}}%

\newcommand{\ii}{\mathrm{i}}%

\newcommand{\dif}{\mathrm{d}}%
\newcommand{\Nabla}{\vec{\nabla}}%
\newcommand{\uu}{\hat{u}}

\newcommand{\Tr}{\operatorname{Tr}}%
\newcommand{\ZT}[1]{\textquotedblleft#1\textquotedblright}%
\newcommand{\Tensor}[1]{\underline{\boldsymbol{#1}}}

\newcolumntype{Y}{>{\centering\arraybackslash}X}%
\newcolumntype{Z}{>{\raggedright\arraybackslash}X}%

\newlength{\myl}%
\newcommand{\SUM}[2]{{\setlength{\myl}{\widthof{$\displaystyle\sum_{#1}^{#2}$}*\real{0.5}-\widthof{$\displaystyle\sum$}*\real{0.5}}\sum_{#1}^{#2}\;\hspace{-\the\myl}}}
\newcommand{\INT}[3]{\settowidth{\myl}{$\displaystyle\int_{#1}^{#2}$}{\int_{#1}^{#2}\;\;\;\hspace{-\the\myl}\dif #3}\,}
\newcommand{\TINT}[3]{\settowidth{\myl}{$\int_{#1}^{#2}$}{\int_{#1}^{#2}\!\ifthenelse{\equal{#1#2}{}}{}{\;\;\;\;\hspace{-\the\myl}}\dif #3}\,}%
\newcommand{\EINT}[3]{\settowidth{\myl}{$\int_{#1}^{#2}$}{\int_{#1}^{#2}\;\;\;\,\hspace{-\the\myl}\dif #3}\,}

\newcommand{\generalizedvelocity}{\vec{\mathfrak{v}}}
\newcommand{\generalizedangularvelocity}{\mathfrak{w}}

\begin{document}
\title{Microscopic field theory for active Brownian particles with translational and rotational inertia}

\author{Michael te Vrugt}
\email[Corresponding author: ]{tevrugtm@uni-mainz.de}
\affiliation{Institut f\"ur Physik, Johannes Gutenberg-Universit\"at Mainz, 55128 Mainz, Germany}

\begin{abstract}
While active matter physics has traditionally focused on particles with overdamped dynamics, recent years have seen an increase of experimental and theoretical work on active systems with inertia. This also leads to an increased need for theoretical models that describe inertial active dynamics. Here, we present a microscopic derivation for a general continuum model describing the nonequilibrium thermodynamics of inertial active matter that generalizes several previously existing works. It applies to particles with translational and rotational inertia and contains particle density, velocity, angular velocity, temperature, polarization, velocity polarization, and angular velocity polarization as dynamical variables. We moreover discuss to which extend commonly used approximations (factorization and local equilibrium) used in the derivation of hydrodynamic models are applicable to inertial active matter.
\end{abstract}
\maketitle

\section{Introduction}
The physics of active matter \cite{BechingerdLLRVV2016,MarchettiJRLPRS2013,teVrugtW2025,teVrugtLC2025} has grown into one of the most flourishing areas of soft matter physics. A key reason for this fact are the unusual thermodynamic properties of active systems, such as the fact that they allow to extract work from a heat bath \cite{SokolovAGA2010} or that one observes here liquid-gas phase separation in purely repulsive systems \cite{CatesT2015}. For the later effect in particular, field-theoretical models \cite{CatesN2025} have proven to be an extremely useful tool.

While many traditionally studied active matter systems exhibit overdamped dynamics, by now also active particles with inertia are studied more and more \cite{Loewen2020,ScholzJLL2018,AntonovCLSL2024,LuoLYP2025,MayerW2022,CapriniM2021b,FengO2025,OmarEtAl2023}. An example for inertial active particles that has particular practical importance are robotic active systems, which tend to be inertial when they are macroscopic \cite{LiPWKWAL2024,LiPDCWDAL2025}. Moreover, a variety of recent works have studied possible realizations of quantum-mechanical active matter \cite{KhassehWMWH2023,NadolnyBB2025}. These systems are often based on ultracold atoms \cite{ZhengL2023,AntonovEtAl2025} and thus very far from having overdamped dynamics. Also from a thermodynamic perspective inertial active fluids are of particular interest, as again they exhibit a number of unexpected phenomena. Among these is the existence of spontaneously developing temperature gradients between coexisting phases \cite{MandalLL2019,HechtDL2024}, the existence of third-order (jerky) dynamics \cite{Lowen2025,teVrugtJW2021}, and unusual cooling phenomena \cite{AntonovMLC2025,MayoEtAl2025}. All of this has motivated also a number of works developing field-theoretical models for inertial active systems \cite{AroldS2020,AroldS2020b,teVrugtJW2021,teVrugtFHHTW2023}. A particularly general model of this kind, studying also temperature differences and velocity correlations, was developed in Ref.\ \cite{MarconiPC2021}. It has been found that in certain limits, such models can be mapped onto the Madelung equations \cite{Madelung1927,VanF2006}, which constitute a hydrodynamic representation of the Schr\"odinger equation \cite{teVrugtFHHTW2023}.

In this work, we employ the interaction-expansion method \cite{BickmannW2020,teVrugtBW2022} to derive a microscopic field theory for active particles with translational and rotational inertia that goes beyond existing models of this type \cite{teVrugtFHHTW2023} by incorporating a larger set of dynamical variables. It describes the coupled dynamics of particle density, velocity, angular velocity, temperature, polarization, velocity polarization, and angular velocity polarization. The relevance of the latter two is a particularly notable feature of the nonequilibrium thermodynamics of inertial active matter. We anticipate our results to be of relevance for future studies of collective dynamics in inertial active matter. In particular, it makes continuum modeling more easily applicable to the case of systems with rotational inertia \cite{CapriniGL2022,Sandoval2020}. Moreover, we provide theoretical insights into the conditions under which some commonly used approximations can be applied. 

\section{Derivation}
\subsection{Microscopic model}
We describe a two-dimensional system of $N$ underdamped active Brownian particles using the Langevin equations \cite{Loewen2020}
		\begin{align}
		\dot{\vec{r}}_i &= \frac{\vec{p}_i}{m},\label{langevin1}\\    
		\dot{\vec{p}}_i &= - \gamma \vec{p}_i - \Nabla_{\vec{r}_i} U(\{\vec{r}_i,\varphi_i\}) + m\gamma v_0 \uu_i + \vec{\eta}_i\label{langevin2},\\
		\dot{\varphi}_i &=\frac{l_i}{I} \label{langevin3},\\
        \dot{l}_i &= - \gamma_\mathrm{R}\varphi_i - \partial_{\varphi_i}U(\{\vec{r}_i,\varphi_i\})+M+ \chi_i\label{langevin4}
		\end{align}
		Here, $\vec{r}_i(t)$, $\vec{p}_i(t)$, $\varphi_i(t)$, and $l_i(t)$ are position, momentum, orientation, and angular momentum of the $i$-th particle, $\uu_i(\varphi_i) = (\cos(\varphi_i),\sin(\varphi_i))^\mathrm{T}$ is the orientation vector (which we assume to specify the direction of the self-propulsion force), $M$ the active torque, $m$ its mass, $I$ its angular momentum, $v_0$ its self-propulsion velocity, $\gamma$ the translational friction coefficient, $\gamma_\mathrm{R}$ the rotational friction coefficient, and $U = U_2 + U_1$ the potential with interaction potential $U_2$ and external potential $U_1$. Moreover, $\vec{\eta}_i$ and $\chi_i(t)$ are translational and rotational noises assumed to have the properties
		\begin{align}
		\braket{\vec{\eta}_i(t)}&=\vec{0},\label{noise1}\\
		\braket{\vec{\eta}_i(t)\otimes\vec{\eta}_j(t')}&=2\gamma m k_{\mathrm{B}} T_{\mathrm{b}}\Eins\delta_{ij}\delta(t-t'),\label{noise2}\\
		\braket{\chi_i(t)}&=0,\label{noise3}\\
		\braket{\chi_i(t)\chi_j(t')}&= 2\gamma_\mathrm{R} I k_{\mathrm{B}} T_{\mathrm{b}}\delta_{ij}\delta(t-t'),\label{noise4}
		\end{align}
		with ensemble average $\braket{\cdot}$, unit matrix $\Eins$, bath temperature $T_\mathrm{b}$, and Boltzmann constant $k_\mathrm{B}$. 

The Fokker-Planck equation corresponding to \cref{langevin1,langevin2,langevin3,langevin4} is given by
		\begin{equation}
		\dot{P}_N(\{\vec{r}_i,\vec{p}_i,\uu_i,l_i\}) = \ii L (\{\vec{r}_i,\vec{p}_i,\uu_i,l_i\})P_N(\{\vec{r}_i,\vec{p}_i,\uu_i,l_i\})\label{fokkerplanck}     
		\end{equation}
		with the $N$-body density $P_N$ and the Liouvillian
		\begin{align}
		&\ii L (\{\vec{r}_i,\vec{p}_i,\uu_i,l_i\})
		\nonumber\\&=\sum_{i=1}^{N}\bigg(-\frac{\vec{p}_i}{m}\cdot \Nabla_{\vec{r}_i} + \gamma+\gamma\vec{p}_i\cdot\Nabla_{\vec{p}_i}+ (\Nabla_{\vec{r}_i}U)\cdot \Nabla_{\vec{p}_i} 
		\nonumber\\&- m\gamma v_0 \uu_i\cdot\Nabla_{\vec{p}_i}+ \gamma m k_{\mathrm{B}} T_\mathrm{b}\Nabla_{\vec{p}_i}^2\nonumber\\ 
  &-\frac{l_i}{I}\partial_{\varphi_i} + \gamma_\mathrm{R}+\gamma_\mathrm{R}l_i\partial_{l_i}+ (\partial_{\varphi_i}U)\partial_{l_i}
		\nonumber\\&- M\partial_{l_i}+ \gamma_\mathrm{R} I k_{\mathrm{B}} T_\mathrm{b}\partial_{l_i}^2\bigg).
		\end{align}
We now integrate \cref{fokkerplanck} over the coordinates of all particles except for one, giving
  	\begin{align}
		&\dot{P}_1(\vec{r},\vec{p},\uu,l)
		\nonumber\\&=\bigg(-\frac{\vec{p}}{m}\cdot\Nabla + \gamma + \gamma\vec{p}\cdot\Nabla_{\vec{p}}+(\Nabla U_1) \cdot\Nabla_{\vec{p}}
		\label{f1}\\& - m\gamma v_0 \uu\cdot\Nabla_{\vec{p}} + \gamma m k_{\mathrm{B}} T \Nabla_{\vec{p}}^2 \nonumber\\
 & -\frac{l}{I}\partial_\varphi + \gamma_\mathrm{R} + \gamma_\mathrm{R}l\partial_l+(\partial_\varphi U_1) \partial_l\nonumber
		\\& - M\partial_l + \gamma_\mathrm{R} I k_{\mathrm{B}} T \partial_l^2
  \bigg)P_1(\vec{r},\vec{p},\uu,l)
		\nonumber\\&+\INT{}{}{^2r_2}\INT{}{}{^2p_2}\INT{0}{2\pi}{\varphi_2}\INT{}{}{l_2}
		\nonumber\\
 &( (\Nabla U_2)\cdot\Nabla_{\vec{p}}+(\partial_\varphi U_2)\partial_l)P_2(\vec{r},\vec{r}_2,\vec{p},\vec{p}_2,\uu,\uu_2,l,l_2). \nonumber
		\end{align}
		The index 1 has been dropped for all coordinates.
  We have introduced the $n$-body density \cite{Archer2009}
		\begin{align}
		\nonumber P_{n}&=\frac{N!}{(N-n)!}\INT{}{}{^2 r_1}\dotsb\INT{}{}{^2 r_{N-n}}\INT{}{}{^2 p_1}\dotsb\INT{}{}{^2 p_{N-n}}
		\\&\INT{0}{2\pi}{\varphi_1}\dotsb\INT{0}{2\pi}{\varphi_{N-n}}\INT{}{}{l_1}\dotsb\INT{}{}{l_{N-n}}P_N
		\end{align}
		with $n\in\{1,\dotsc,N\}$.

\subsection{Factorization approximation}
In the case of passive simple fluids, where only positions and momenta of the particle are relevant, one would now usually make the factorization assumption \cite{MarconiM2010}
\begin{equation}
P_2(\vec{r},\vec{r}_2,\vec{p},\vec{p}_2)\approx P_1(\vec{r},\vec{p})P_1(\vec{r}_2,\vec{p}_2)g(\vec{r},\vec{r}_2)
\label{factor}
\end{equation}
with the pair-distribution function $g(\vec{r},\vec{r}_2)$. The function $P_1$ is then further approximated by the local equilibrium form \cite{Archer2009}
\begin{equation}
P_1(\vec{r},\vec{p})=\frac{\rho(\vec{r})}{2mk_\mathrm{B}T}\exp\bigg(-\frac{(\vec{p}-m\vec{v}(\vec{r},t)^2}{2mk_\mathrm{B}T}\bigg)
\label{localeq}
\end{equation}
with a velocity field $\vec{v}$. In Ref.\ \cite{teVrugtFHHTW2023} (in the step from Eq.\ (64) to Eq.\ (69)), the authors generalized this by allowing $g$ and $\vec{v}$ to also depend on particle orientations (which is crucial for motility-induced phase separation \cite{BialkeLS2013}), but still implicitly assumed that the velocity dependence can be factorized in this way. However, it has been found that the velocities of different particles in inertial active matter exhibit significant correlations \cite{CapriniM2021}. Such correlations (which also exist for overdamped active Brownian particles \cite{CapriniMP2020}) are further enhanced by the presence of rotational inertia \cite{CapriniGL2022}. It may seem as if the existence of such correlations prevents one from making an approximation of the form \eqref{factor}. However, as we will demonstrate now, such an approximation is in fact still possible, such that the ansatz used in Ref.\ \cite{teVrugtFHHTW2023} is valid.

Velocity correlations emerge if the expectation value $\braket{\vec{p}\cdot\vec{p}'}(\vec{r},\vec{r}')$ with the momenta $\vec{p}$ and $\vec{p}'$ of two different particles located at $\vec{r}$ and $\vec{r}'$ does not vanish. Let us calculate this expectation value for a system with a two-body distribution given by \cref{factor,localeq}:
\begin{equation}
\begin{split}
&\braket{\vec{p}\cdot\vec{p}'}(\vec{r},\vec{r}') \\&=\INT{}{}{^2p}\INT{}{}{^2p'}\vec{p}\cdot\vec{p}'P_1(\vec{r},\vec{p})P_1(\vec{r}_2,\vec{p}_2)g(\vec{r},\vec{r}')\\  
 &= \vec{v}(\vec{r},t)\cdot\vec{v}(\vec{r}',t)g(\vec{r},\vec{r}').
\end{split}    
\end{equation}
Consequently, when there is a non-vanishing velocity field $\vec{v}$, then there are velocity correlations (except in the very special case $g=0$). Intuitively, this makes a lot of sense. When the velocities of nearby particles are correlated, then nearby particles move a similar direction, which corresponds to an overall net flow. The simplest example is a system where all particles move in the same direction. In this case, the velocities of nearby particles would be very strongly correlated, and the hydrodynamic manifestation of this would be the existence of a non-vanishing (and in this case homogeneous) velocity field $\vec{v}$. An isolated passive system with friction will eventually settle to a state with $\vec{v}=\vec{0}$, i.e., without velocity correlations. The key observation of Ref.\ \cite{CapriniMP2020} was that this is not the case for active systems.

Moreover, it can be expected that the orientation-dependent pair-distribution function $g(\vec{r},\vec{r}',\hat{u},\hat{u}')$, which was obtained numerically in Refs.\ \cite{JeggleSW2020,BrokertVJSW2024}, in principle contains all information required to obtain the velocity correlations. (Assuming that at least for weak inertia they look somewhat similar, this result carries over to the inertial case.) For overdamped particles, the momentum $\vec{p}$ is a function of $\vec{r}$ and $\hat{u}$. (While the momentum is determined by a stochastic equation, one can derive a \ZT{deterministic velocity} that also does the job for calculating averaged quantities \cite{SchmidtB2013}.) We can thus easily calculate the velocity correlation as
\begin{align}
&\braket{\vec{p}\cdot\vec{p}'}(\vec{r},\vec{r}') \\&=\INT{0}{2\pi}{\varphi}\INT{0}{2\pi}{\varphi'}\vec{p}(\vec{r},\hat{u})\cdot\vec{p}(\vec{r}',\hat{u}')\varrho(\vec{r},\hat{u})\varrho(\vec{r}',\hat{u}')g(\vec{r},\vec{r}',\hat{u},\hat{u}').\notag
\end{align}

\subsection{Generalized local equilibrium approximation}
A further approximation that plays a key role in microscopic derivations of hydrodynamic equations is the local equilibrium approximation \cite{Archer2009}. It can be generally motivated from the projection operator formalism \cite{Grabert1982,teVrugt2022}: Provided that we have a set of relevant variables $A_i$ with given mean values enforced by Lagrange multipliers $\lambda_i$, a suitable ansatz for the phase-space distribution $\bar{\rho}$ is
\begin{equation}
\bar{\rho}(t)=\frac{1}{Z(t)}e^{-\lambda_i A_i},
\end{equation}
where the partition function $Z$ ensures a normalization. In this work, generalizing the result from Ref.\ \cite{teVrugtFHHTW2023}, we make a local equilibrium approximation of the form
\begin{equation}
\begin{split}
&P_{\mathrm{leq}}(\vec{r},\vec{p},\uu,l)\\=&\frac{\varrho(\vec{r},\uu)}{(2\pi k_{\mathrm{B}} T(\vec{r},\uu))^\frac{3}{2} m \sqrt{I}}\\&\exp\!\bigg(-\beta(\vec{r},\uu)\bigg(\frac{(\vec{p}-m\generalizedvelocity(\vec{r},\uu))^2}{2m}+\frac{(l-I\generalizedangularvelocity(\vec{r},\uu))^2}{2I}\bigg)\bigg)    
\end{split}
\label{localeq1}
\end{equation}   
with the local thermodynamic beta
$\beta = (k_\mathrm{B}T)^{-1}$, the generalized velocity $\generalizedvelocity$, and the generalized angular velocity $\generalizedangularvelocity$. All these quantities are \textit{defined} via \cref{localeq1} (see Ref.\ \cite{teVrugtFHHTW2023} for a discussion of this point). We assume the two-body density $P_2$ to have the form
\begin{equation}
\begin{split}
&P_2(\vec{r},\vec{r}_2,\vec{p},\vec{p}_2,\uu,\uu_2,l,l_2) \\&= P_{\mathrm{leq}}(\vec{r},\vec{p},\uu,l)P_{\mathrm{leq}}(\vec{r}_2,\vec{p}_2,\uu_2,l_2)g(\vec{r},\vec{r}_2,\uu,\uu_2),    
\end{split}
\label{fachere}
\end{equation}
where we moreover assume the pair-distribution function $g$ to be time-independent. This generalizes \cref{factor} towards the active case. We have introduced the one-body density
\begin{equation}
\varrho(\vec{r},\uu)=\INT{}{}{^2p}\INT{}{}{l}P_1(\vec{r},\vec{p},\uu,l).\label{obd}
\end{equation}
\cref{f1,localeq1,fachere,obd} give, by integrating over $\vec{p}$ and $l$, the result
\begin{equation}
\dot{\varrho} = - \Nabla\cdot(\varrho\generalizedvelocity) -\partial_\varphi (\varrho\generalizedangularvelocity).
\label{dotvarrho}
\end{equation}
Similarly, multiplying \cref{f1} by $\vec{p}$ and integrating out gives
\begin{equation}
\begin{split}
&\partial_t(\varrho\generalizedvelocity) + \Nabla\cdot(\varrho \generalizedvelocity\otimes\generalizedvelocity)+\frac{1}{m}\Nabla(k_B T \varrho)\\=& - \gamma \varrho\generalizedvelocity + \gamma v_0 \varrho \uu -\frac{1}{m}\varrho\Nabla U_1 -\frac{1}{m}\vec{\mathcal{I}}_{\mathrm{trans}}
\end{split}
\label{dotgeneralizedv}
\end{equation}
with the translational interaction term
\begin{equation}
\vec{\mathcal{I}}_{\mathrm{trans}}=\INT{}{}{^3r_3}\INT{0}{2\pi}{\varphi_2}\varrho(\vec{r},\uu)\varrho(\vec{r}_2,\uu_2) g(\vec{r},\vec{r}_2,\uu,\uu_2)\Nabla U_2,
\label{transinteractionterm}
\end{equation}
having used the fact that \cref{localeq1} implies
\begin{equation}
\varrho(\vec{r},\uu)\generalizedvelocity(\vec{r},\uu) = \INT{}{}{^2p}\INT{}{}{l}\frac{\vec{p}}{m}P_1(\vec{r},\vec{p},\uu,l).
\end{equation}  
Now we enter new territory compared to previous work: We multiply \cref{f1} by $l$ and integrate out, giving
\begin{equation}
\begin{split}
&\partial_t(\varrho\generalizedangularvelocity) + \partial_\varphi(\varrho \generalizedangularvelocity^2)+\frac{1}{I}\partial_\varphi(k_B T \varrho)\\=& - \gamma_\mathrm{R} \varrho\generalizedangularvelocity -\frac{1}{I}\varrho\partial_\varphi U_1 + \frac{M}{I}\varrho -\frac{1}{I}\mathcal{I}_{\mathrm{rot}}
\end{split}
\label{dotgeneralizedangularvel}
\end{equation}
with the rotational interaction term
\begin{equation}
\mathcal{I}_{\mathrm{rot}}=\INT{}{}{^3r_3}\INT{0}{2\pi}{\varphi_2}\varrho(\vec{r},\uu)\varrho(\vec{r}_2,\uu_2) g(\vec{r},\vec{r}_2,\uu,\uu_2)\partial_\varphi U_2
\end{equation}
having used the fact that \cref{localeq1} implies
\begin{equation}
\varrho(\vec{r},\uu)\generalizedangularvelocity(\vec{r},\uu) = \INT{}{}{^2p}\INT{}{}{l}\frac{l}{I}P_1(\vec{r},\vec{p},\uu,l).
\end{equation} 
Finally, we multiply \cref{f1} by $\vec{p}^2/(2m)+l^2/(2I)$ and find
\begin{equation}
\begin{split}
&\partial_t \bigg(\frac{3}{2}k_B T + \frac{1}{2}m\generalizedvelocity^2 + \frac{1}{2}I \generalizedangularvelocity^2\bigg)\varrho\\=&   - \generalizedvelocity\cdot (\Nabla U_1 - m\gamma v_0 \uu +\vec{\mathcal{I}}_{\mathrm{trans}}) \\&- \generalizedangularvelocity(\partial_\phi U_1 - M +\mathcal{I}_{\mathrm{rot}})\\&+\gamma \bigg(k_\mathrm{B}T_\mathrm{b}-k_\mathrm{B}T-\frac{1}{2}m\generalizedvelocity^2\bigg)\varrho\\&+\gamma_\mathrm{R}\bigg(k_\mathrm{B}T_\mathrm{b}-k_\mathrm{B}T -\frac{1}{2}I\generalizedangularvelocity^2\bigg)\varrho,
\end{split}
\label{dott}
\end{equation}
having used the fact that \cref{localeq1} implies
\begin{equation}
\begin{split}
&\bigg(\frac{3}{2}k_B T + \frac{1}{2}m\vec{\generalizedvelocity}^2 + \frac{1}{2}I \generalizedangularvelocity^2\bigg)\varrho \\=& \INT{}{}{^2p}\INT{}{}{l}\bigg(\frac{\vec{p}^2}{2m}+ \frac{l^2}{2I}\bigg)P_1(\vec{r},\vec{p},\uu,l).
\end{split}
\end{equation} 
\subsection{Approximations of interaction terms}
Following previous work \cite{BickmannW2020,teVrugtFHHTW2023}, we assume that $g$ is translationally and rotationally invariant and can therefore be parametrized as $g(r,\theta_1,\theta_2)$ with $\theta_1 = \varphi_{\mathrm{R}}-\varphi$ and $\theta_2 = \varphi_2 -\varphi$ and the parametrization $\vec{r}_2-\vec{r} = r\uu(\varphi_{\mathrm{R}})$. A combined Fourier and gradient expansion gives \cite{BickmannW2020,teVrugtFHHTW2023}
\begin{equation}
\begin{split}
\vec{\mathcal{I}}_\mathrm{trans}&=-\sum_{l=0}^{\infty}\frac{1}{l!}\varrho(\vec{r},\uu,t)\INT{0}{\infty}{r}r^{l+1}\INT{0}{2\pi}{\varphi_{\mathrm{R}}}\uu(\varphi_{\mathrm{R}})(\uu(\varphi_{\mathrm{R}})\cdot\Nabla)^l \\&\INT{0}{2\pi}{\varphi_2}\sum_{n_1,n_2 = -\infty}^{\infty} g_{\mathrm{trans},n_1 n_2}(r)\cos(n_1\theta_1+n_2\theta_2)\varrho(\vec{r},\uu_2,t),   
	\end{split}
\label{transinteractions}
		\end{equation}
where the coefficients are given by 
		\begin{equation}
  \begin{split}
		&g_{\mathrm{trans},n_1 n_2}(r) \\=& \frac{\TINT{0}{2\pi}{\theta_1}\TINT{0}{2\pi}{\theta_2}\, U_2'(r, \theta_1, \theta_2)g(r, \theta_1, \theta_2)\cos(n_1\theta_1+n_2\theta_2)}{4\pi^2}      
  \end{split}
		\end{equation}
		and the prime indicates a derivative with respect to $r$. We have assumed here that $U_2$ is a function of $r$, $\theta_1$, and $\theta_2$.

Similarly, we make a Fourier and gradient expansion for the rotational interaction term, which gives
\begin{equation}
\begin{split}
\mathcal{I}_\mathrm{rot}&=\sum_{l=0}^{\infty}\frac{1}{l!}\varrho(\vec{r},\uu,t)\INT{0}{\infty}{r}r^{l+1}\INT{0}{2\pi}{\varphi_{\mathrm{R}}}(\uu(\varphi_{\mathrm{R}})\cdot\Nabla)^l \\&\INT{0}{2\pi}{\varphi_2}\sum_{n_1,n_2 = -\infty}^{\infty} g_{\mathrm{rot},n_1 n_2}(r)\cos(n_1\theta_1+n_2\theta_2)\varrho(\vec{r},\uu_2,t),   
	\end{split}
\label{rotinteractions}
\end{equation}
  with the coefficients
  \begin{equation}
  \begin{split}
		&g_{\mathrm{rot},n_1 n_2}(r) \\=& \frac{\TINT{0}{2\pi}{\theta_1}\TINT{0}{2\pi}{\theta_2}\, (\partial_\varphi U_2(r, \theta_1, \theta_2))g(r, \theta_1, \theta_2)\cos(n_1\theta_1+n_2\theta_2)}{4\pi^2}.      
  \end{split}
  \end{equation}
 The essential difference to previous works in this step is the angular dependence of $U_2$, requiring us to Fourier expand the products $U_2'g$ and $(\partial_\varphi U_2)g$ rather than just $g$, and the fact that we have two interaction terms rather than just one. Note that the sign difference between \cref{transinteractionterm,transinteractions} arises from the way we have defined $\hat{u}(\varphi_\mathrm{R})$. Since this vector does not appear in \cref{rotinteractions}, this equation has no minus sign.

 \subsection{Orientational expansions}
As a next step, as in Ref.\ \cite{teVrugtFHHTW2023} (but now with a larger number of fields, we carry out the Cartesian orientational expansions \cite{teVrugtW2020b}
		\begin{align}
		\varrho(\vec{r},\uu)&=\rho(\vec{r})+\uu\cdot\vec{P}(\vec{r}),\label{expansion1}\\
		\generalizedvelocity(\vec{r},\uu)&=\vec{v}(\vec{r})+\uu\cdot \Tensor{v}_{\vec{P}}(\vec{r}),\label{expansion2}\\
  		\generalizedangularvelocity(\vec{r},\uu)&=\omega(\vec{r})+\uu\cdot\vec{W}(\vec{r}),\\
    U_1(\vec{r},\uu) &= \bar{U}_{1}(\vec{r}) + \uu \cdot \vec{u}_1(\vec{r})
		\end{align}
		with the non-orientational particle density 
		\begin{equation}
		\rho(\vec{r})=\frac{1}{2\pi}\INT{0}{2\pi}{\varphi}\varrho(\vec{r},\uu),   
		\end{equation}
		local velocity
		\begin{equation}
		\vec{v}(\vec{r})=\frac{1}{2\pi}\INT{0}{2\pi}{\varphi}\generalizedvelocity(\vec{r},\uu),    
		\end{equation}
        local angular velocity
		\begin{equation}
		\omega(\vec{r})=\frac{1}{2\pi}\INT{0}{2\pi}{\varphi}\generalizedangularvelocity(\vec{r},\uu),   
		\end{equation}     
  effective temperature
  		\begin{equation}
		T_{\mathrm{eff}}(\vec{r})=\frac{1}{2\pi}\INT{0}{2\pi}{\varphi}T(\vec{r},\uu),   
		\end{equation}     
  orientation-averaged external potential
  		\begin{equation}
		\bar{U}_1(\vec{r})=\frac{1}{2\pi}\INT{0}{2\pi}{\varphi}U_1(\vec{r},\uu),   
		\end{equation}
		local polarization
		\begin{equation}
		\vec{P}(\vec{r})=\frac{1}{\pi}\INT{0}{2\pi}{\varphi}\uu\varrho(\vec{r},\uu),
		\end{equation}
		local velocity polarization 
		\begin{equation}
		\Tensor{v}_{\vec{P}}(\vec{r})=\frac{1}{\pi}\INT{0}{2\pi}{\varphi}\uu\otimes\generalizedvelocity(\vec{r},\uu),   
		\end{equation}
local angular velocity polarization
  		\begin{equation}
		\vec{W}(\vec{r})=\frac{1}{\pi}\INT{0}{2\pi}{\varphi}\uu\generalizedangularvelocity(\vec{r},\uu),
		\end{equation}
and external potential polarization
		\begin{equation}
		\vec{u}_1(\vec{r})=\frac{1}{\pi}\INT{0}{2\pi}{\varphi}\uu U_1(\vec{r},\uu).
		\end{equation}
As discussed in Ref.\ \cite{teVrugtFHHTW2023}, the necessity to incorporate observables like $\Tensor{v}_{\vec{P}}$ and $\vec{W}$ arises from the fact that in the generalized local equilibrium approximation \eqref{localeq1} we allow the fields $\generalizedvelocity$ and $\generalizedangularvelocity$ to depend on both position and orientation. This, in turn, is necessary to correctly capture phenomena like motility-induced phase separation in the overdamped limit, as otherwise we could not derive a density-dependent swimming speed. Note that for the temperature, we work only with the orientational average as otherwise the dynamics would be too complex.

A further reason for the importance of such variables are the observations made in Ref.\ \cite{MandalLL2019}: Defining the kinetic temperature as
\begin{equation}
T_\mathrm{kin} = \frac{1}{2mk_\mathrm{B}}\braket{\vec{p}^2},
\end{equation}
one finds a difference in kinetic temperature between gas and dense phase of a phase-separating system as
\begin{equation}
T_{\mathrm{gas}}- T_{\mathrm{dense}} = \frac{v_0}{2k_\mathrm{B}}(\braket{\vec{p}\cdot\hat{u}}_{\mathrm{gas}}-\braket{\vec{p}\cdot\hat{u}}_{\mathrm{dense}}).
\end{equation}
Physically speaking, temperature differences arise from the fact that momentum and orientation of an inertial active particle are correlated in the dilute and uncorrelated in the dense phase. Such correlations are, however, precisely what $\Tensor{v}_{\vec{P}}$, being the first orientational moment of the velocity distribution, captures. Consequently, field-theoretical models for motility-induced temperature differences will require this observable.

With these expansions, \cref{transinteractions} becomes
\begin{equation}
\begin{split}
\INT{0}{2\pi}{\varphi}\vec{\mathcal{I}}_{\mathrm{trans}}&= A_1\vec{P}\rho + A_2\Nabla \rho^2 + A_3\Nabla\vec{P}^2+A_4 \vec{P}\Nabla^2\rho \\&+2A_4(\vec{P}\cdot\Nabla)\Nabla\rho + A_5\rho\Nabla^2\vec{P}+2A_5\rho\Nabla(\Nabla\cdot\vec{P})
\end{split}
\label{expandeditrans}
\end{equation}
with
\begin{align}
A_1&= -2\pi^3 \INT{0}{\infty}{r}r (g_{\mathrm{trans},10}(r)+g_{\mathrm{trans},-10}(r)\notag\\&\quad +g_{\mathrm{trans},1-1}(r)+g_{\mathrm{trans},-11}(r)),\\
A_2&= -2\pi^3 \INT{0}{\infty}{r}r^2 g_{\mathrm{trans},00}(r),\\
A_3&=-\frac{\pi^3}{2}\INT{0}{\infty}{r}r^2 (g_{\mathrm{trans},01}(r)+g_{\mathrm{trans},0-1}(r)),\\
A_4&=-\frac{\pi^3}{4}\INT{0}{\infty}{r}r^3 (g_{\mathrm{trans},10}(r)+g_{\mathrm{trans},-10}(r)),\\
A_5&=-\frac{\pi^3}{4}\INT{0}{\infty}{r}r^3 (g_{\mathrm{trans},1-1}(r)+g_{\mathrm{trans},-11}(r))
\end{align}
and (truncating the gradient expansion at $l=0$)
\begin{equation}
\begin{split}
\INT{0}{2\pi}{\varphi}\uu\otimes\vec{\mathcal{I}}_{\mathrm{trans}}&= A_6\rho^2 \Eins +A_7 \vec{P}\otimes\vec{P} + A_8 \vec{P}^2\Eins
\end{split}
\label{expandeditrans2}
\end{equation}
with
\begin{align}
A_6&=2\pi^3\INT{0}{\infty}{r}r(g_{10}(r)+g_{-10}(r)),\\
A_7&=\pi^3\INT{0}{\infty}{r}r(g_{1-1}(r)+g_{-11}(r)),\\
A_8&=\frac{\pi^3}{2}\INT{0}{\infty}{r}r(g_{11}(r)+g_{-1-1}(r)).
\end{align}
Similarly, from \cref{rotinteractions} we find (truncating the gradient expansion at $l=2$)
\begin{equation}
\begin{split}
\INT{0}{2\pi}{\varphi}\mathcal{I}_{\mathrm{rot}}&= B_1\rho^2 + B_2 \vec{P}^2+B_3 \vec{P}\cdot\Nabla\rho + B_4 \rho \Nabla\cdot\vec{P}\\&+B_5 \vec{P}\cdot(\Nabla^2\vec{P})+B_6\rho\Nabla^2\rho
\end{split}
\label{expandedirot}
\end{equation}
with
\begin{align}
B_1&= 8\pi^3 \INT{0}{\infty}{r}r g_{\mathrm{rot},00}(r),\\
B_2&= 2\pi^3 \INT{0}{\infty}{r}r (g_{\mathrm{rot},01}(r)+g_{\mathrm{rot},0-1}(r)),\\
B_3&=2\pi^3 \INT{0}{\infty}{r}r^2 (g_{\mathrm{rot},10}(r)+g_{\mathrm{rot},-10}(r)),\\
B_4&=2\pi^3 \INT{0}{\infty}{r}r^2 (g_{\mathrm{rot},1-1}(r)+g_{\mathrm{rot},-11}(r)),\\
B_5&=\frac{\pi^3}{2} \INT{0}{\infty}{r}r^3 (g_{\mathrm{rot},01}(r)+g_{\mathrm{rot},0-1}(r)),\\
B_6&=2\pi^3 \INT{0}{\infty}{r}r^3 g_{\mathrm{rot},00}(r)
\end{align}
and (truncating the gradient expansion at $l=0$)
\begin{equation}
\INT{0}{2\pi}{\varphi}\uu\mathcal{I}_{\mathrm{rot}}= B_7 \vec{P}\rho
\label{expandedirot2}
\end{equation}
with
\begin{equation}
B_7=2\pi^3\INT{0}{\infty}{r}r(2g_{00}(r)+g_{01}(r)+g_{0-1}(r)).
\end{equation}

From \cref{dotvarrho}, we find
\begin{align}
\partial_t \rho &= -\Nabla\cdot\bigg(\rho \vec{v}+ \frac{1}{2}\vec{P}\cdot\Tensor{v}
_{\vec{P}}\bigg),\\
\partial_t \vec{P}&= -\Nabla\cdot(\rho\Tensor{v}
_{\vec{P}}+ \vec{v}\otimes\vec{P}) + \Tensor{\epsilon}\cdot(\omega \vec{P} + \rho \vec{W})
\end{align}
with the Levi-Civita tensor
\begin{equation}
\Tensor{\epsilon}=
\begin{pmatrix}
0 & 1\\
-1 & 0
\end{pmatrix}.
\end{equation}
From \cref{dotgeneralizedv}, we get
\begin{align}
\begin{split}
&\partial_t \bigg(\rho \vec{v}+ \frac{1}{2}\vec{P}\cdot\Tensor{v}
_{\vec{P}}\bigg)\\
&+\Nabla\cdot\bigg(\rho\vec{v}\otimes\vec{v}+\frac{1}{2}(\rho \Tensor{v}
_{\vec{P}}\cdot \Tensor{v}
_{\vec{P}}+(\vec{v}\otimes\vec{P})\cdot\Tensor{v}
_{\vec{P}}\\
 &+(\vec{P}\cdot\Tensor{v}
_{\vec{P}})\otimes\vec{v})\bigg)+\frac{k_\mathrm{B}}{m}\Nabla(\rho T_\mathrm{eff})\\
 =&-\gamma \bigg(\rho \vec{v}+ \frac{1}{2}\vec{P}\cdot\Tensor{v}
_{\vec{P}}\bigg)\\
& - \frac{1}{m}\bigg(\rho \Nabla \bar{U}_1 + \frac{1}{2}(\Nabla \vec{u}_1)\cdot\vec{P}\bigg)\\
& -\frac{1}{m}(A_1\vec{P}\rho + A_2\Nabla \rho^2 + A_3\Nabla\vec{P}^2+A_4 \vec{P}\Nabla^2\rho \\&+2A_4(\vec{P}\cdot\Nabla)\Nabla\rho + A_5\rho\Nabla^2\vec{P}+2A_5\rho\Nabla(\Nabla\cdot\vec{P})),
\end{split}
\label{dottv}
\end{align}
having used \cref{expandeditrans}, and
\begin{equation}
\begin{split}
&\partial_t(\rho\Tensor{v}_{\vec{P}}+\vec{P}\otimes\vec{v}) + \Nabla \cdot(\rho\vec{v}\otimes \Tensor{v}_{\vec{P}})+ \Nabla \cdot(\rho\Tensor{v}_{\vec{P}}\otimes \vec{v})\\
&+\Nabla\cdot(\vec{v}\otimes\vec{P}\otimes\vec{v})+ \frac{1}{4}\Nabla\otimes\vec{P}:\Tensor{v}_{\vec{P}} \Tensor{v}_{\vec{P}} \\&+ \frac{1}{4}\Nabla\cdot(\Tensor{v}_{\vec{P}}\cdot\Tensor{v}_{\vec{P}}\otimes\vec{P}) + \frac{1}{4}\Nabla\cdot(\Tensor{v}_{\vec{P}} \vec{P}\cdot\Tensor{v}_{\vec{P}})\\&+ \frac{1}{m}\Nabla\otimes(k_\mathrm{B}T_{\mathrm{eff}}\vec{P})\\
=& -\gamma (\rho \Tensor{v}_{\vec{P}} + \vec{P}\otimes\vec{v}) + \gamma v_0 \rho \Eins - \frac{1}{m} \vec{P}\otimes \Nabla \bar{U}_1 \\&- \frac{1}{m}\rho (\Nabla \otimes \vec{u}_1)^\mathrm{T}\\
& - \frac{1}{m\pi}(A_6\rho^2 \Eins +A_7 \vec{P}\otimes\vec{P} + A_8 \vec{P}^2\Eins),
\end{split}
\label{dotvp}
\end{equation}
having used \cref{expandeditrans2}. For clarity, we also show the latter result in index notation:
\begin{equation}
\begin{split}
&\partial_t(\rho V_{ij} + P_i v_j) + \partial_m (\rho v_m V_{ij}) + \partial_m (\rho V_{mi}v_j) \\
& + \partial_m (v_m  P_i v_j) + \frac{1}{4}\partial_m (P_n V_{mn}V_{ij}) + \frac{1}{4}\partial_m (P_i V_{mn}V_{nj}) \\&+ \frac{1}{4}\partial_m (P_n V_{mi}V_{nj}) + \frac{1}{m}\partial_i (k_\mathrm{B}T_{\mathrm{eff}}P_j)\\
=& - \gamma(\rho V_{ij} + P_i v_j) +  \gamma v_0 \rho \delta_{ij} - \frac{1}{m}P_i \partial_j \bar{U}_1 - \frac{1}{m}\rho \partial_j u_{1i}\\
&- \frac{1}{m\pi}(A_6\rho^2\delta_{ij} + A_7 P_i P_j + A_8 P_m^2\delta_{ij}).
\end{split}
\end{equation}
Here, $V_{ij}$ are the components of the tensor $\Tensor{v}_{\vec{P}}$. Note that \cref{dottv,dotvp} have a slightly different form than Eqs. (89) and (90) in Ref.\ \cite{teVrugtFHHTW2023} since in the present work we did not divide by $\varrho$ before making the expansions.

From \cref{dotgeneralizedangularvel}, we get
\begin{equation}
\begin{split}
&\partial_t \bigg(\rho \omega + \frac{1}{2}\vec{P}\cdot\vec{W}\bigg) \\&= \frac{M}{I}\rho - \gamma_\mathrm{R} \bigg(\rho \omega + \frac{1}{2}\vec{P}\cdot\vec{W}\bigg) - \frac{1}{I}\vec{P}\cdot\Tensor{\epsilon}\cdot\vec{u}_1\\
&-\frac{1}{2\pi I}(B_1\rho^2 + B_2 \vec{P}^2+B_3 \vec{P}\cdot\Nabla\rho + B_4 \rho \Nabla\cdot\vec{P}\\&+B_5 \vec{P}\cdot(\Nabla^2\vec{P})+B_6\rho\Nabla^2\rho),
\end{split}
\end{equation}
having used \cref{expandedirot}, and
\begin{equation}
\begin{split}
&\partial_t (\omega\vec{P} + \rho\vec{W}) + 2\omega\rho \Tensor{\epsilon}\cdot\vec{W} + \frac{1}{4}\vec{W}^2\Tensor{\epsilon}\cdot\vec{P} \\&+ \frac{1}{2}\vec{P}\cdot\vec{W}\Tensor{\epsilon}\cdot\vec{W} +\omega^2\Tensor{\epsilon}\cdot\vec{P}+ \frac{1}{I}T_{\mathrm{eff}} \Tensor{\epsilon}\cdot\vec{P}\\
&=\frac{M}{I} \vec{P} - \gamma_\mathrm{R}(\omega \vec{P} + \rho\vec{W})- \frac{1}{I}\rho\Tensor{\epsilon}\cdot\vec{u}_1 - \frac{B_7}{\pi I} \vec{P}\rho,
\end{split}
\end{equation}
having used \cref{expandedirot2}.

Finally, \cref{dott} becomes
\begin{equation}
\begin{split}
&\partial_t\bigg(\frac{3}{2}k_\mathrm{B}T_{\mathrm{eff}}\rho + \frac{1}{2}m\rho\vec{v}^2+\frac{1}{2}m\vec{P}\cdot\Tensor{v}_{\vec{P}}\cdot\vec{v}+\frac{1}{4}m\rho\Tensor{v}_{\vec{P}}:\Tensor{v}_{\vec{P}} \\&+ \frac{1}{2}I\omega^2\rho + \frac{1}{2}I \omega \vec{P}\cdot\vec{W}+ \frac{1}{4}I \rho\vec{W}^2\bigg)\\
&-\vec{v}\cdot\Nabla\bigg(\bar{U}_1 - \frac{1}{2}\Tensor{v}_{\vec{P}}:(\Nabla\otimes\vec{u}_1) +\frac{1}{2}\gamma m v_0 \Tr(\Tensor{v}_{\vec{P}})\\&+ \frac{1}{2}\vec{u}_1\cdot\Tensor{\epsilon}\cdot\vec{W} + M\omega\bigg)\\
&-\vec{v}\cdot(A_1\vec{P}\rho + A_2\Nabla \rho^2 + A_3\Nabla\vec{P}^2+A_4 \vec{P}\Nabla^2\rho \\&+2A_4(\vec{P}\cdot\Nabla)\Nabla\rho + A_5\rho\Nabla^2\vec{P}+2A_5\rho\Nabla(\Nabla\cdot\vec{P}))\\
&-\omega(B_1\rho^2 + B_2 \vec{P}^2+B_3 \vec{P}\cdot\Nabla\rho + B_4 \rho \Nabla\cdot\vec{P}\\&+B_5 \vec{P}\cdot(\Nabla^2\vec{P})+B_6\rho\Nabla^2\rho)\\
& + \gamma \bigg(k_\mathrm{B}T_\mathrm{b}- k_\mathrm{B}T_{\mathrm{eff}}-\frac{1}{2} m \rho\vec{v}^2 - \frac{1}{4} m \Tensor{v}_{\vec{P}}:\Tensor{v}_{\vec{P}}\bigg)\\
&+ \gamma_\mathrm{R} \bigg(k_\mathrm{B}T_\mathrm{b}- k_\mathrm{B}T_{\mathrm{eff}}-\frac{1}{2}I \omega^2 - \frac{1}{2}I \omega \vec{P}\cdot\vec{W} - \frac{1}{4} I \rho \vec{W}^2\bigg).
\end{split}
\end{equation}
We have neglected products of $\Tensor{v}_{\vec{P}}$ and $\vec{W}$ with the interaction terms in this equation and used \cref{expandeditrans,expandedirot}. Of particular physical interest is the term $\gamma m v_0 \Tr(\Tensor{v}_{\vec{P}})/2$, which indicates (as discussed above) that in the case of an inertial active system the field $\Tensor{v}_{\vec{P}}$ is responsible for temperature differences.

\section{Conclusion}
In this work, we have developed a general continuum model for active particles with translational and rotational inertia. It contains as dynamical variables the zeroth and first angular moment of density, velocity, and angular velocity as well as the local kinetic temperature. We have derived the model from the microscopic Langevin equations using the interaction-expansion method. 

Clearly, the resulting model is incredibly complex and therefore of somewhat limited practical use. An obvious next step would therefore be to consider limiting cases in which only several of these variables are considered, but in which certain effects of interest still survive. This would then enable a variety of interesting applications to, e.g., robotic active systems, active dusty plasmas, or quantum active matter. Moreover, our results provide insights into a number of interesting theoretical questions, in particular concerning the question which approximations (local equilibrium ansatz, factorization approximation etc.) are (not) permissible in the case of inertial active systems. Such questions do even in the overdamped case often turn out to be more complex than often assumed \cite{BurekovicDLCN2026,KalzSM2024,BoltzKI2024}.

\acknowledgments{I would like to thank Lorenzo Caprini and Sangyun Lee for seminal discussions. This work is funded by the Deutsche Forschungsgemeinschaft (DFG, German Research Foundation) -- Project-ID 525063330.}

%

\end{document}